\documentclass[showpacs,preprintnumbers,amsmath,amssymb]{revtex4}

\usepackage{epsfig}
\usepackage{graphicx}
\usepackage{dcolumn}
\usepackage{bm}
\usepackage{epstopdf}
\usepackage{subfigure}

\newcommand{\EQ}{\begin{equation}}
\newcommand{\EN}{\end{equation}}
\newcommand{\ea}{\end{eqnarray}}
\newcommand{\ba}{\begin{eqnarray}}

\newcommand{\bear}{\begin{eqnarray}}
\newcommand{\ear}{\end{eqnarray}}

\begin{document}


\title{Elementary objects of the 1D Hubbard model}
\author{J. M. P. Carmelo}
\affiliation{Department of Physics and Center of Physics, University of Minho, Campus Gualtar, P-4710-057 Braga, Portugal}
\affiliation{Beijing Computational Science Research Center, Beijing 100084, China}
\affiliation{Institut f\"ur Theoretische Physik III, Universit\"at Stuttgart, D-70550 Stuttgart, Germany}


\begin{abstract}
Exotic elementary objects such as ``holons'' and ``spinons'', which are widely used in descriptions of 
correlated electrons in reduced spatial dimensions, were introduced from analysis of the 
excitation branches of one-dimensional (1D) models. The 1D Hubbard model with effective 
nearest-neighbor hopping integral $t$ and on-site repulsion $U$ is a prominent example.
In the last twenty years a large number of angle-resolved photoemission spectroscopy 
experiments as well as electron energy-loss spectroscopic studies and high-resolution resonant 
inelastic X-ray experiments on several quasi-1D metals and quasi-1D Mott-Hubbard insulators have 
observed separate charge and spin spectral features, which have been identified with ``holons'' and 
``spinons''. The elementary objects emerging within non-perturbative 1D correlated systems play now the same role in actual low-dimensional 
materials as Fermi-liquid quasiparticles in three-dimensional metals. The 1D Hubbard model  
elementary-object representations belong to two groups: those whose elementary-object occupancy 
configurations generate the energy eigenstates from (i) the electron or hole vacuum 
and (ii) from a given ground state. Most of the holon and spinon representations belong to the latter class. 
Alternative representations as those in terms of ``pseudoparticles'', ``pseudofermions'', and
on-site repulsion $U\rightarrow\infty$ spineless fermions and Heisenberg spins refer to the former group. 
For the study of the model elementary excitations and static properties, both
types of representations correspond to equivalent ``dual pictures''. However,
the model dynamical correlations involve scattering events of the elementary
objects of type (i) that are not described within the scattering theories of those of type (ii). The main
goal of this paper is the review of representations of the 1D Hubbard model physics
in terms of elementary objects whose configurations generate the energy
eigenstates from the electron or hole vacuum. In addition, the relation to the holon and 
spinon representations is discussed and clarified. The introduction of the several 1D Hubbard model
elementary-object representations followed a mere association 
of such objects with Bethe-ansatz (BA) solution quantum numbers, often along with symmetry arguments. 
The relation of such elementary objects operator algebra to the electron creation and annihilation operators remained a
longstanding open issue. The studies of this paper profit from the interplay of the exact BA solution with its 
recently found extended global $[SU(2)\otimes SU(2)\otimes U(1)]/Z_2^2$ symmetry to unify the results 
being reviewed. This includes to clarify issues related to the elementary-object operator algebra and its 
non-perturbative relation to the electron operators. The main microscopic mechanism is an
electron - rotated-electron unitary transformation performed by the BA solution. 
All alternative 1D Hubbard model elementary-object
representations revisited in this review are found to be related to 
three basic elementary objects that emerge from the separation of 
the rotated-electron occupancy configurations degrees of freedom. 
The occupancy configurations of such three basic
elementary objects generate both all the model energy eigenstates and 
corresponding independent state representations of the group
$[SU(2)\otimes SU(2)\otimes U(1)]/Z_2^2$ associated with its global symmetry.
Particular attention is paid to the review of the model pseudoparticle and pseudofermion representations.
The $c$ pseudoparticles are shown to be one of the three basic elementary objects
that emerge from the rotated electrons degrees of freedom separation. On the
other hand, it is confirmed that all pseudoparticles belonging to the remaining branches are $\eta$-spin-neutral or
spin-neutral composite objects of an even number of $\eta$-spin-$1/2$ or spin-$1/2$
basic elementary objects, respectively. The corresponding pseudofermion
scattering and dynamical theories are shortly reviewed and the
consequences of the relation to the model global symmetry clarified. Furthermore,
the relation of two alternative holon and spinon representations to the three basic elementary
objects that emerge from the rotated electrons is clarified. The different elementary-object
representations are found to refer to alternative sets of degenerate energy
eigenstates that span well-defined model subspaces. The role of the
elementary objects in the properties of the 1D Hubbard model
confirms that the spectral features observed in actual photoemission
experiments on quasi-1D materials refer to this review elementary objects.
\end{abstract}


\maketitle

\section{Contents}
\label{contents}

The final version of the present review paper whose list of contents is given in the following
will be posted soon in the arXiv.\\ \\ 
1. Introduction

1.1 Exotic elementary objects in low-dimensional electronic correlated systems and quasi-1D materials

1.2 The ``simplest'' paradigm for correlated electrons on a 1D lattice: The Hubbard model

1.3 Isomorphism between symmetry group state representations and energy eigenstates

1.4 The thermodynamic BA equations in functional form

1.5 Elementary-object representations relative to the electron or hole vacuum

1.6 Elementary-object representations relative to a ground state\\
2. Rotated electrons and corresponding three basic elementary objects

2.1 General rotated electrons from the model global symmetry alone

2.2 The specific rotated electrons arising from global symmetry and the BA solution

2.3 Three basic elementary objects emerging from the rotated electrons degrees of freedom separation\\
3. Three effective lattices and the spinon and $\eta$-spinon transformation laws

3.1 Rotated-electron wave-function factorization and degrees of freedom separation

3.2 Three effective lattices

3.3 Consequences of the spinon and $\eta$-spinon transformation laws

3.4 The absolute $\eta$-spinon and spinon vacua and the $\eta$-spinon and spinon subspace\\
4. The $\eta$-spin-neutral and spin-neutral configurations

4.1 Set of $\nu=1,...,\infty$ $\eta$-spin-singlet $2\nu$-$\eta$-spinon and spin-singlet $2\nu$-spinon configurations

4.2 Subspace dependent $\alpha\nu$ bond particle statistics

4.3 The $\alpha\nu$ pseudoparticles emerging from the hard-core $\alpha\nu$ bond particles

4.4 The symmetry algebra representations generated by the elementary-object occupancy configurations\\
5. The $\beta$ pseudoparticle, unbound spinon, and unbound $\eta$-spinon representation

5.1 The $\beta$ pseudoparticle energy functional

5.2 Unbound $\eta$-spinon and unbound spinon energies and towers of elementary current quanta

5.3 Anti-binding $\eta$-spinon ($\alpha =\eta$) and binding spinon 
($\alpha =s$) energy and $\nu>1$ pairs interaction

5.4 $\alpha\nu$ pseudoparticle transformation laws under the electron - rotated-electron unitary transformation\\
6. The 1D Hubbard model pseudofermion representation

6.1 The $\beta$-pseudoparticle - $\beta$-pseudofermion unitary transformation

6.2 The $\beta$ pseudofermion non-interacting energy functional and exotic operator algebra

6.3 Pseudofermion scattering theory

6.4 The shadow-host and shadow-neutral mechanisms

6.5 Charge and spin stiffnesses at finite and zero temperature

6.6 Pseudofermion dynamical theory

6.7 Relation to the low-energy conformal-field spectrum and two-component Tomonaga-Luttinger liquid\\
7. Relation to other elementary-object representations and their scattering theories

7.1 The $\beta$ pseudofermion and the two most used holon-spinon representations

7.2 Relation between the three elementary-object representations scattering theories

7.3 Different elementary-object representations refer to different choices of degenerate energy eigenstates\\
8. Outlook

8.1 Elementary objects associated with the rotated-electron wave-function factorization

8.2 Dynamical and transport properties as controlled by the elementary-objects microscopic processes

8.3 Future developments\\
Acknowledgements\\
Appendix A. Two-pseudofermion phase shifts\\
Appendix B. The pseudofermion  energy spectrum\\
References

\end{document}